\PassOptionsToPackage{table}{xcolor}
\documentclass[sigconf]{acmart}

\usepackage{xcolor}
\definecolor{lightgray}{gray}{0.9}
\usepackage{amstext}
\usepackage{amsmath}
\usepackage{enumitem}
\usepackage[show]{chato-notes}
\usepackage{algorithm}
\usepackage{xspace}
\usepackage{algpseudocode}
\usepackage{dblfloatfix}

\newcommand{\idnt}{\phantom{ - }}
\newcommand{\sys}[1]{\textsc{Gar}\def\temp{#1}\ifx\temp\empty{}\else\raisebox{-.4ex}{\scriptsize #1}\fi}
\newcommand{\sysbm}[1]{\sys{}${}_{BM25}$}
\newcommand{\crc}[1]{#1}

\newcommand{\gy}[1]{\textcolor{gray}{#1}}
\newcommand{\nic}[1]{\textcolor{black}{#1}}
\newcommand{\crf}[1]{\textcolor{black}{#1}}

\usepackage{marginnote}
\newcommand{\pageenlarge}[1]{\enlargethispage{#1\baselineskip}}

\definecolor{MyGreen}{HTML}{548235}
\def\HiLi{\leavevmode\rlap{\hbox to 26em{\color{MyGreen!25}\leaders\hrule height .8\baselineskip depth .5ex\hfill}}}
\def\HiLii{\leavevmode\rlap{\hbox to 24.5em{\color{MyGreen!25}\leaders\hrule height .8\baselineskip depth .5ex\hfill}}}
\def\HiLiii{\leavevmode\rlap{\hbox to 24.5em{\color{MyGreen!25}\leaders\hrule height 1.6\baselineskip depth 3.0ex\hfill}}}

\AtBeginDocument{%
  \providecommand\BibTeX{{%
    \normalfont B\kern-0.5em{\scshape i\kern-0.25em b}\kern-0.8em\TeX}}}

\copyrightyear{2022} 
\acmYear{2022} 
\setcopyright{acmcopyright}\acmConference[CIKM '22]{Proceedings of the 31st ACM International Conference on Information and Knowledge Management}{October 17--21, 2022}{Atlanta, GA, USA}
\acmBooktitle{Proceedings of the 31st ACM International Conference on Information and Knowledge Management (CIKM '22), October 17--21, 2022, Atlanta, GA, USA}
\acmPrice{15.00}
\acmDOI{10.1145/3511808.3557231}
\acmISBN{978-1-4503-9236-5/22/10}

\begin{document}

\title{Adaptive Re-Ranking with a Corpus Graph}

\author{Sean MacAvaney}
\email{sean.macavaney@glasgow.ac.uk}
\affiliation{%
  \institution{University of Glasgow}
  \city{Glasgow}
  \country{United Kingdom}
}

\author{Nicola Tonellotto}
\email{nicola.tonellotto@unipi.it}
\affiliation{%
  \institution{University of Pisa}
  \city{Pisa}
  \country{Italy}
}

\author{Craig Macdonald}
\email{craig.macdonald@glasgow.ac.uk}
\affiliation{%
  \institution{University of Glasgow}
  \city{Glasgow}
  \country{United Kingdom}
}

\renewcommand{\shortauthors}{MacAvaney, et al.}

\begin{abstract}
\looseness -1 Search systems often employ a re-ranking pipeline, wherein documents (or passages) from an initial pool of candidates are assigned new ranking scores. The process enables the use of highly-effective but expensive scoring functions that are not suitable for use directly in structures like inverted indices or approximate nearest neighbour indices. However, re-ranking pipelines are inherently limited by the recall of the initial candidate pool; documents that are not identified as candidates for re-ranking by the initial retrieval function cannot be \crc{identified}. We propose a novel approach for overcoming the recall limitation based on the well-established clustering hypothesis.
Throughout the re-ranking process, our approach adds documents to the pool that are most similar to the highest-scoring documents up to that point. This feedback process \textit{adapts} the pool of candidates to those that may also yield high ranking scores, even if they were not present in the initial pool. It can also \nic{increase the score of} documents that appear deeper in the pool that would have otherwise been skipped due to a limited re-ranking budget. We find that our \textit{Graph-based Adaptive Re-ranking} (\sys{}) approach significantly improves the performance of re-ranking pipelines in terms of precision- and recall-oriented measures, is complementary to a variety of existing techniques (e.g., dense retrieval), is robust to its hyperparameters, and contributes minimally to computational and storage costs. \nic{For instance, on the MS~MARCO passage ranking dataset, \sys{} can improve the nDCG of a BM25 candidate pool by up to 8\%  when applying a monoT5 ranker.\footnote{\crc{Code for this work is available at \url{https://github.com/terrierteam/pyterrier_adaptive}.}}}

\end{abstract}



\maketitle

\algdef{SE}[DOWHILE]{Do}{doWhile}{\algorithmicdo}[1]{\algorithmicwhile\ #1}%

\section{Introduction}
\looseness -1 Deep neural ranking models -- especially those that use contextualised language models like BERT~\cite{bert} -- have brought significant benefits in  retrieval effectiveness across a range of tasks~\cite{lin2020pretrained}. %
The most effective techniques tend to be those that first retrieve a pool of candidate documents\footnote{Or passages; we often simply use the term ``document'' for ease of reading.} using an inexpensive retrieval approach and then re-score them using a more expensive function. This process is called \textit{re-ranking}, since the documents from the candidate pool are given a new ranked order. Re-ranking enables the use of sophisticated scoring functions (such as cross-encoders, which jointly model the texts of the query and document) that are incompatible with \nic{inverted indexes} or \nic{vector} indexes. Since the scoring function can be computationally expensive, re-ranking is often limited to a predefined maximum number documents that the system is willing to re-rank for each query (i.e., a \textit{re-ranking budget}, such as 100).

The performance of a re-ranking pipeline is limited by the recall of the candidate pool, however. This is because documents that were not found by the initial ranking function have no chance of being re-ranked. Consequently, a variety of techniques are employed to improve the recall of the initial ranking pool, including document-rewriting approaches that add semantically-similar terms to an inverted index~\cite{docTTTTTquery}, or dense document retrieval techniques that enable semantic searching~\cite{colbert}.

\begin{figure}
\centering
\includegraphics[scale=0.45]{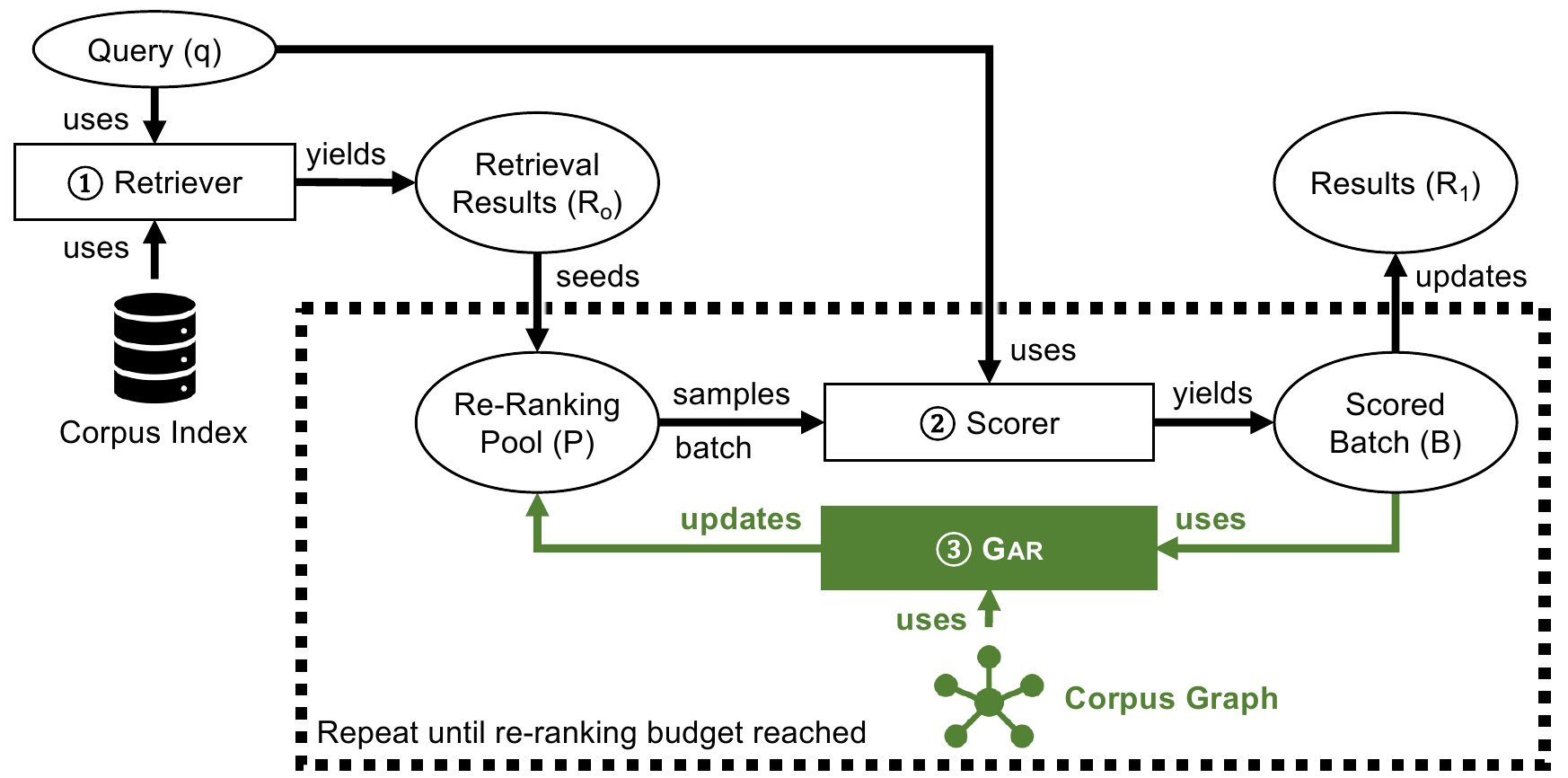}\vspace{-0.5em}
\caption{Overview of \textsc{Gar}. Traditional re-ranking exclusively scores results seeded by the retriever. \textsc{Gar} (in green) adapts the re-ranking pool after each batch based on the computed scores and a pre-computed graph of the corpus.}
\label{fig:overview}\vspace{-1em}
\end{figure}

\looseness -1 In this work, we explore a complementary approach to overcoming the recall limitation of re-ranking based on the long-standing \textit{clustering hypothesis}~\cite{jardine-1971-use}, which suggests that closely-related documents tend to be relevant to the same queries. \nic{During the re-ranking process, our approach, \nic{called} \textit{Graph-based Adaptive Re-Ranking} (\sys{}), prioritises the scoring of the neighbours of documents that have received high scores up to this point.}
An overview of \sys{} is shown in Figure~\ref{fig:overview}. The \sys{} feedback mechanism allows for documents to be retrieved that were not present in the initial re-ranking pool, which can improve system recall. It also allows for the re-ranking of the documents that may have otherwise been skipped from the pool when the re-ranking budget is low. Finally, by including the feedback within the re-ranking itself (as opposed to post-scoring feedback mechanisms, \nic{such as PRF}), our approach can find documents that are multiple hops away (i.e., neighbours of neighbours). \sys{} achieves low online overhead \nic{through offline computation of} a \textit{corpus graph} that stores the nearest neighbours of each document.

On experiments over the TREC Deep Learning datasets, we find that \sys{} significantly improves precision- and recall-oriented evaluation measures. \sys{} can improve virtually any re-ranking pipeline, with the results largely holding across a variety of initial retrieval functions (lexical, dense retrieval, document expansion, and learned lexical), scoring functions (cross-encoding, late interaction),
\nic{document} similarity \nic{metrics}
(lexical, semantic), and re-ranking budgets (high, low). Impressively, a \sys{} pipeline that uses only BM25 for \nic{both the initial retrieval and the document similarity} is able to achieve comparable or improved performance in terms of re-ranked precision and recall over the competitive TCT-ColBERT-v2-HNP~\cite{lin-etal-2021-batch} and DocT5Query~\cite{docTTTTTquery} models -- both of which \crc{have} far \crc{higher} requirements in terms of offline computation and/or storage \nic{capacities}. We find that the online overhead of \sys{} is low compared to a typical re-ranking, usually only adding around 2-4ms per 100 documents re-ranked. We also find that \sys{} is largely robust to its parameters, with major deviations in performance only occurring \crc{with extreme parameter values}. Finally, we find that despite using document similarity, \sys{} does not significantly reduce the diversity among the relevant retrieved documents.

\looseness -1 \crf{In summary, we propose a novel approach to embed a feedback loop within the neural re-ranking process to help identify un-retrieved relevant documents through application of the clustering hypothesis.} \nic{Our contributions can therefore be summarised as follows: (1) We demonstrate a novel application of the clustering hypothesis in the context of neutral re-ranking; (2) We show that our proposed approach can successfully improve both the precision and the recall of re-ranking pipelines with minimal computational overhead; (3) We demonstrate that the approach is robust across pipeline components and the parameters it introduces.}
The remainder of the paper is organised as follows: We first provide additional background and related work, positioning \sys{} in context with past work in neural retrieval, relevance feedback, and the clustering hypothesis (Section~\ref{sec:background}); We then briefly demonstrate that the clustering hypothesis still holds on a recent dataset to motivate our approach (Section~\ref{sec:initial_analysis}); We formally describe our method (Section~\ref{sec:meth}) and present our experiments that demonstrate its effectiveness (Sections~\ref{sec:exp} \&~\ref{sec:res}); We wrap up with final conclusions and future directions of this promising area~(Section~\ref{sec:concl}).

\section{Background and Related Work}\label{sec:background}
The recent advancements in deep neural ranking models have brought significant improvements on the effectiveness of ad-hoc ranking tasks in IR system~\cite{lin2020pretrained}. In particular, pre-trained language models such as BERT~\cite{bert} and T5~\cite{t5} are able to lean semantic representations of words depending on their context, and these representations are able to better model the relevance of a document w.r.t. a query, with notable improvements w.r.t. classical approaches. 
However, these improvements have an high computational costs; BERT-based rankers~\cite{bertir,cedr} are reported to be slower than classical rankers such as those based on BM25 by orders of magnitude~\cite{https://doi.org/10.48550/arxiv.1907.04614,cedr}. Therefore, it is still usually infeasible to directly use pre-trained language models to rank all documents in a corpus for each query \crc{(even using various to reduce the computational cost~\cite{macavaney:sigir2020-epic,colbert,macavaney:sigir2020-eff}.)} Deep neural ranking models are typically deployed as re-rankers in a pipeline architecture, where a first preliminary ranking stage is deployed before the more expensive neural re-ranker, in a cascading manner. During query processing, the first ranking stage retrieves from the whole document corpus a candidate pool of documents using a simple ranking function, with the goal of maximising the recall effectiveness. The following re-ranking stage processes the documents in the candidate pool, reordering them by focusing on high precision results at the top positions, whose documents will be returned to the user~\cite{fntir,monobert}.
In this setting, there is an efficiency-effectiveness tradeoff on the number of documents retrieved by the first ranker. From the efficiency perspective, a smaller number of documents in the candidate pool will allow the re-ranker to reduce the time spent on re-ranking the documents, since the execution time is proportional to the candidate set size. From the effectiveness perspective, the larger the candidate pool, the higher the number of potentially relevant documents to be retrieved from the document corpus. In fact, relevant documents can be retrieved from the corpus only during first-stage processing. The recall effectiveness of the candidate pool has been investigated in previous IR settings, in particular in learning-to-rank pipelines. \citet{10.1145/2433396.2433407} studied how, given a time budget, dynamic pruning strategies~\cite{fntir} can be use in first-stage retrieval to improve the candidate pool size on a per-query basis. \citet{MacDonald2012TheWA} studied the minimum effective size of the document pool, i.e., when to stop ranking in the first stage, and concluded that the smallest effective pool for a given query depends, among others, on the type of the information need and the document representation.
\nic{In the context of neural IR, learned sparse retrieval focuses on learning new terms to be included in a document before indexing, and the impact scores to be stored in the inverted index, such that the resulting ranking function approximates the effectiveness of a full transformer-based ranker while retaining the efficiency of the fastest inverted-index based methods~\cite{https://doi.org/10.48550/arxiv.2205.04733,deepimpactv1,deepct}. In doing so, first-stage rankers based on learned impacts are able to improve the recall w.r.t. BM25, but the end-to-end recall is still limited by the first-stage ranker.}

Pseudo-Relevance Feedback (PRF) involves the reformulation of a query based on the top results (e.g., by adding distinctive terms from the top documents). This query is then re-issued to the engine, producing a new ranked result list. Adaptive Re-Ranking also makes use of these top-scoring documents, but differs in two important ways. First, the query remains unmodified, and therefore, ranking scores from the model need not be re-computed. Second, the top scores are used in an intermediate stage of the scoring process; the process is guided by the highest-scoring documents known up until a given point, which may not reflect the overall top results. Finally, we note that the output of an adaptive re-ranking operation could be fed as input into a PRF operation to perform query reformulation.

This work can be seen as a modern instantiation of the clustering hypothesis, which \citet{jardine-1971-use} stated as {\em ``Closely associated documents tend to be relevant to the same requests"}. Many works have explored the clustering hypothesis for various tasks in information retrieval, such as for visualisation of the corpus (e.g.,~\cite{10.1145/502585.502592}), visualisation of search results (e.g.,~\cite{10.1145/133160.133214}), enriching document representations~\cite{10.1145/1008992.1009027} and fusing rankings (e.g.,~\cite{10.1145/1390334.1390428}). Most related to our application is the usage of the clustering hypothesis for first-stage retrieval (i.e., document selection), in which the documents to rank are identified by finding the most suitable cluster for a query~\cite{kurland-thesis}. \nic{However, these works focus on identifying the most suitable clusters for a given query and transforming the constituents into a ranking. Moreover, while our approach also takes a soft clustering approach~\cite{10.1145/2484028.2484192} where each `cluster' is represented by a document and its neighbours, instead of ranking clusters, we identify ``good'' clusters as when the representative document is scored highly by a strong neural scoring function. We also address the problem of transforming the documents into a ranking by letting the neural scoring function do that job as well.} \crf{Overall, our novel approach is the first to embed a feedback loop within the re-ranking process to help identify un-retrieved relevant documents.}

\section{Preliminary Analysis}\label{sec:initial_analysis}

{
\renewcommand{\arraystretch}{0.8}
\begin{table}[b]
\centering
\vspace{-1em}
\caption{Distribution of nearest \crc{neighbouring passages}, among pairs of judged passages in TREC DL 2019, based on BM25 and TCT-ColBERT-HNP similarity scores. Each cell represents the percentage that a passage with a given relevance label ($x$) has a nearest neighbour with the column's relevance label ($y$); each row sums to 100\%.}\vspace{-0.75em}
\label{tab:motiv_dist}
\begin{tabular}{lcccc}
\multicolumn{5}{c}{\bf BM25}\\
\toprule
&\multicolumn{4}{c}{neighbour's rel $y$}\\
  & 0 & 1 & 2 & 3 \\
\cmidrule(lr){2-5}
\multicolumn{1}{r|}{$x=0$} &\bf67 &   11 &   16 &    7 \\
\multicolumn{1}{r|}{1}     &   14 &\bf47 &   31 &    8 \\
\multicolumn{1}{r|}{2}     &    8 &   12 &\bf71 &    9 \\
\multicolumn{1}{r|}{3}     &    8 &    7 &   12 &\bf73 \\
\bottomrule
\end{tabular}
\hspace{1em}
\begin{tabular}{lcccc}
\multicolumn{5}{c}{\bf TCT-ColBERT-HNP}\\
\toprule
&\multicolumn{4}{c}{neighbour's rel $y$}\\
  & 0 & 1 & 2 & 3 \\
\cmidrule(lr){2-5}
\multicolumn{1}{r|}{$x=0$} &\bf76 &   10 &   10 &    4 \\
\multicolumn{1}{r|}{1}     &   17 &\bf46 &   29 &    8 \\
\multicolumn{1}{r|}{2}     &    8 &   11 &\bf72 &    8 \\
\multicolumn{1}{r|}{3}     &    6 &    7 &   12 &\bf75 \\
\bottomrule
\end{tabular}
\end{table}
}

\pageenlarge{1}We first perform a preliminary check to see whether the clustering hypothesis appears to hold on a recent dataset and using a recent model. Namely, we want to check whether the passages from the MS~MARCO corpus~\cite{Bajaj2016Msmarco} are more likely to distributed closer to those with the same relevance labels than those with differing grades. We explore two techniques for measuring similarity: a lexical similarity score via BM25, and a semantic similarity via TCT-ColBERT-HNP~\cite{lin-etal-2021-batch}. For the queries in the TREC DL 2019 dataset~\cite{10.1145/3404835.3463249}, we compute similarity scores between each pair of judged documents. Then, \crc{akin to the Voorhees' cluster hypothesis test~\cite{10.1145/253495.253524},} we calculate the distribution of the relevance labels of the nearest \crc{neighbouring passage} by relevance label (i.e., we calculate $P\big(rel(neighbour(p))=y|rel(p)=x\big)$ for all pairs of relevance labels $x$ and $y$.)

Table~\ref{tab:motiv_dist} presents the results of this analysis. We observe a clear trend: passages with a given relevance label are far more likely to be \nic{closer to the passages with the same label (among judged passages) than those with other labels (in the same row)}. This holds across both lexical (BM25) and semantic (TCT-ColBERT) similarity measures, and across all four relevance labels (ranging from non-relevant to perfectly relevant).

This analysis suggests that the clustering hypothesis holds on TREC DL 2019. Therefore, it follows that the neighbours of passages that a scoring function considers most relevant are a reasonable place to look for additional relevant passages to be scored -- which is the core motivation of our proposed method.

\section{Graph-based Adaptive Re-Ranking}\label{sec:meth}

\algdef{SE}[DOWHILE]{Do}{doWhile}{\algorithmicdo}[1]{\algorithmicwhile\ #1}%

\begin{figure}[t]\vspace{-1em}
\begin{algorithm}[H]
\caption{Graph-based Adaptive Re-Ranking}\label{alg:arr}
\begin{algorithmic}
\Require Initial ranking $R_0$, batch size $b$, budget $c$, corpus graph $G$
\Ensure Re-Ranked pool $R_1$
\State $R_1 \gets \emptyset$ \Comment{Re-Ranking results}
\State\HiLi $P \gets R_0$ \Comment{Re-ranking pool}
\State\HiLi $F \gets \emptyset$ \Comment{Graph frontier}
\Do
  \State $B \gets$ \Call{Score}{top $b$ from $P$, subject to $c$} \Comment{e.g., monoT5}
  \State $R_1 \gets R_1 \cup B$ \Comment{Add batch to results}
  \State $R_0 \gets R_0 \setminus B$ \Comment{Discard batch from initial ranking}
  \State\HiLii $F \gets F \setminus B$ \Comment{Discard batch from frontier}
  \State\HiLii $F \gets F \cup (\Call{Neighbours}{B, G} \setminus R_1$) \Comment{Update frontier}
  \State\HiLiii $P \gets \begin{cases} 
      R_0 & \text{if}\; P = F   \\
      F   & \text{if}\; P = R_0 \\
   \end{cases}$ \Comment{Alternate initial ranking and frontier}
\doWhile{$|R_1| < c$}
\State $R_1 \gets R_1 \cup \Call{Backfill}{R_0, R_1}$ \Comment{Backfill remaining items}
\end{algorithmic}
\end{algorithm}\vspace{-2em}
\end{figure}

\nic{We now introduce the document re-ranking scenario, and we present a description of our proposed re-ranking algorithm.}
\pageenlarge{1}Let $R_0$ denote an initial ranked pool of $|R_0|$ documents produced by a first-stage ranker, and let $R_1$ denote \nic{a subsequent} re-ranked pool of $|R_1| = |R_0|$ documents. A certain number of top ranked documents from the $R_1$ pool will subsequently be returned to the user who issued the query.
\looseness -1 In re-ranking, we assume that the documents from $R_0$ are processed in batches of $b$ documents at maximum \nic{(the size of the last batch depends on the re-ranking budget).}
A scoring function \textsc{Score()} takes as input a batch of documents, e.g., the top scoring $b$ documents in $R_0$ and re-scores them according to a specific re-ranking stage implementation. The re-scored batch is added the final re-ranked pool $R_1$, and \nic{then} removed from the initial ranked pool $R_0$. Note that by setting $b=1$, we are re-ranking one document at a time, as in classical learning-to-rank scenarios; in contrast, when $b > 1$, we allow for more efficient re-ranking function implementations leveraging advanced hardware, such as GPUs and TPUs.

Since the time available for re-ranking is often small, and given that it is directly proportional to the number of documents re-ranked, the re-ranking process can be provided with a budget $c$, denoting the maximum number of documents to be re-ranked given the proportional time constraint. If the budget does not allow to re-rank all the document in the initial ranked pool, the \textsc{Backfill} function returns the documents in $R_0$ that have not been re-ranked, i.e., not in $R_1$, that are used to fill up the final re-ranked pool $R_1$ to contain all the documents initially included in $R_0$. \nic{For example, if $R_0$ contains 1000 documents and, due to the budget, only 100 documents can be re-scored, the 900 top ranked documents in $R_0$ but not re-ranked in $R_1$ are appended to $R_1$ in the same order as in $R_0$, to obtain a re-ranked list of 1000 documents.} \nic{The uncoloured lines in Alg.~\ref{alg:arr} illustrate this re-ranking algorithm, which corresponds to the common re-ranking adopted in a pipelined cascading architecture.}

In our adaptive re-ranking algorithm, we leverage a corpus graph \nic{$G = (V,E)$}. This \crc{directed} graph encodes the similarity between documents, and can be computed offline, using lexical or semantic similarity function between two documents. Every node in $V$ represents a document in the corpus, and every pair of documents may be connected with an edge in $E$, labelled with the documents' similarity. To address \crc{the graph's} quadratic space (and time) complexity, we limit to a small value $k$ the number of edges for each node in the corpus graph\nic{, i.e., $|E| = k|V|$}. The top $k$ edges are selected according to their similarity scores, in decreasing order.

\looseness -1 Our adaptive re-ranking algorithm, illustrated in Alg.~\ref{alg:arr}, receives  an initial ranking pool of documents $R_0$, a batch size $b$, a budget $c$, and the corpus graph $G$ as input. We consider a dynamically updated re-ranking pool $P$, initialised with the contents of $R_0$ ($P \leftarrow R_0$), and a dynamically updated graph frontier $F$, initially empty ($F \leftarrow \emptyset$). After the re-ranking of the top $b$ documents selected from $P$ and subject to the constraint $c$ \crc{(called batch $B$, where $b=|b|$)}, we update the initial and re-ranked pools $R_0$ and $R_1$. The documents in the batch are removed from the frontier $F$ because there is no need to re-rank them again.
Now we consider the documents in the batch $B$, and we look up in the corpus graph for documents whose nodes are directly connected to the documents in $B$. These documents (except any that have already been scored) are added to the frontier ($F \cup (\textsc{Neighbours}(B, G) \setminus R_1)$), prioritised by the computed ranking score of the source document. Note that the neighbours may occur later in the ranking list. Next, instead of using the current contents of the initial pool $R_0$ for the next batch evaluation, we alternate between $R_0$ and the current frontier $F$. In doing so, \nic{we ensure that $R_1$ contains documents from $R_0$ and newly identified documents not included in $R_0$}. The algorithm proceeds alternating between these two options, populating the frontier at each step, until the budget allows, then backfills the final pool of initial candidates as before.

\pageenlarge{2}\crc{We note that alternating between the initial ranking and the frontier is somewhat na\"ive; perhaps it is better to score more/fewer documents from the frontier, or to dynamically decide whether to select batches from the frontier or the initial ranking based on recent scores. Indeed, we investigated such strategies in pilot experiments but were unable to identify a strategy that consistently performed better than the simple alternating technique. We therefore decided to leave the exploration of alternative techniques to future work.}

\section{Experimental Setup}\label{sec:exp}

We experiment to answer the following research questions:

\begin{enumerate}
\item[\bf RQ1] \nic{What is the impact of \sys\ on retrieval effectiveness compared to typical re-ranking?}
\item[\bf RQ2] What is the computational overhead \nic{introduced by} \sys{}? (Section~\ref{sec:perf})
\item[\bf RQ3] How sensitive is \sys{} to the parameters it introduces: the number of neighbours included in the corpus graph $k$ \nic{and} the batch size $b$? (Section~\ref{sec:robust})
\item[\bf RQ4] \nic{What is the impact of \sys\ on retrieval effectiveness compared to state-of-the-art neural IR systems?}
\end{enumerate}

\noindent Finally, because \sys{} is based on scoring similar documents, we recognise that it has the potential to reduce the diversity of the retrieved passages (i.e., it could make the retrieved passages more homogeneous). Therefore, we ask:

\begin{enumerate}
\item[\bf RQ5] Does \sys{} result in more homogeneous relevant passages than existing techniques?
\end{enumerate}

\subsection{Datasets and Evaluation}

Our primary experiments are conducted using the TREC Deep Learning 2019 (DL19) and 2020 (DL20) test collections~\cite{10.1145/3404835.3463249}. DL19 is used throughout the development and for the analysis of \sys, and therefore acts as our validation set. DL20 is held out until the final evaluation, allowing us to confirm that our approach has not over-fit to DL19. Both datasets use the MS~MARCO passage ranking corpus, which consists of 8.8M passages~\cite{Bajaj2016Msmarco}. DL19 consists of 43 queries and an average of 215 relevance assessments per query; DL20 has 54 queries with 211 assessments per query. We evaluate our approach using nDCG, MAP, and Recall at rank 1000. For the binary measures (MAP and Recall), we use the standard practice of setting a minimum relevance score of 2, which counts answers that are highly or perfectly relevant. \nic{In our experiments} we are concerned with both precision and recall, \nic{so} we focus on nDCG without a rank cutoff, though we also report the official task measure of nDCG with a rank cutoff of 10 (nDCG@10) to provide \nic{meaningful comparisons} with other works. %

\pageenlarge{2}We select DL19 and DL20 because they provide more complete relevance assessments than the MS~MARCO development set\nic{; this} is especially important given that \sys{} is designed to retrieve documents that were not necessarily in the initial re-ranking pool. For completeness, we also report performance on the small subset of MS MARCO dev, which consists of 6980 queries, each with 1.1 relevance assessments per query on average. For this dataset, we report the official measure of Mean Reciprocal Rank at 10 (MRR@10) and the commonly-reported value of Recall at 1000.

\subsection{Retrieval and Scoring Models}

\looseness -1 To test the effect of \sys{} under a variety of initial ranking conditions, we conduct experiments using four retrieval functions \nic{as first stage rankers}, each \nic{representing a different \nic{family of} ranking approaches.}%
\begin{itemize}[leftmargin=*]
\item \textbf{BM25}, a simple and long-standing lexical retrieval approach. We retrieve the top 1000 BM25 results from a PISA~\cite{pisa} index using default parameters.
\item \textbf{TCT}, a dense retrieval approach. We conduct exact (i.e., exhaustive) retrieval of the top 1000 results using a TCT-ColBERT-HNP model~\cite{lin-etal-2021-batch} trained on MS~MARCO.\footnote{Hugging Face ID: \texttt{castorini/tct\_colbert-v2-hnp-msmarco}} This is among the most effective dense retrieval models to date.
\item \textbf{D2Q}, a document expansion approach. We retrieve the top 1000 BM25 results from a PISA index of documents expanded using a docT5query model~\cite{docTTTTTquery} trained on MS~MARCO. We use the expanded documents released by the authors. This is the most effective document expansion model we are aware of to date.
\item \textbf{SPLADE}, a learned \crc{sparse} lexical retrieval model. We retrieve the top 1000 results for a SPLADE++ model~\cite{https://doi.org/10.48550/arxiv.2205.04733} trained on MS~MARCO (\texttt{CoCondenser-EnsembleDistil} version). We use code released by the authors for indexing and retrieval.\footnote{\url{https://github.com/naver/splade}} This is the most effective learned lexical retrieval model we are aware of to date.
\end{itemize}
Similarly, we experiment with the following neural re-ranking models to test the effect of the scoring function on \sys{}.
\begin{itemize}[leftmargin=*]
\item \textbf{MonoT5}, a sequence-to-sequence scoring function. We test two versions of the MonoT5 model~\cite{monot5} trained on MS~MARCO from two base language models: MonoT5-base, and MonoT5-3b. The 3b model has the same structure as the base model, but has more parameters ($13\times$ more; 2.9B, compared to base's 223M) so it is consequently more expensive to run. These models are among the most effective scoring functions reported to date.\footnote{\nic{We also experiment with applying DuotT5~\cite{10.48550/arxiv.2101.05667} as a final re-ranker in Section~\ref{sec:horserace}.}}
\item \textbf{ColBERT} (scorer only), a late interaction scoring function. Although ColBERT~\cite{colbert} can be used in an end-to-end fashion (i.e., using its embeddings to perform dense retrieval), we use it as a scoring function over the aforementioned retrieval functions. The model represents two paradigms: one where representations are pre-computed to reduce the query latency, and another where the \nic{representations} are computed on-the-fly.
\end{itemize}

\noindent \crc{We use the implementations of the above methods provided by PyTerrier~\cite{macdonald:cikm2021-pyterrier}. Following PyTerrier notation,} we use >> to denote a re-ranking pipeline. For instance, ``BM25>>MonoT5-base'' retrieves using BM25 and re-ranks using MonoT5-base.

\subsection{Corpus Graphs}

\looseness -1 \nic{In our experiments, we construct and exploit two corpus graphs, namely a lexical similarity graph and a semantic similarity graph.}
The lexical graph (denoted as \sys{BM25}) is constructed by retrieving the top BM25~\cite{bm25} results using the text of the passage as the query. We use PISA to perform top $k+1$ \nic{lexical} retrieval (discarding the passage itself). Using a 4.0 GHz 24-core AMD Ryzen Threadripper Processor, the MS~MARCO passage graph takes around 8 hours to construct.
The semantic similarity graph (denoted as \sys{TCT}) is constructed using the TCT-ColBERT-HNP model. We perform an exact (i.e., exhaustive) search over an index to retrieve the top $k+1$ most similar embeddings to each passage (discarding the passage itself). Using an NVIDIA GeForce RTX 3090 GPU to compute similarities, the MS~MARCO passage graph takes around 3 hours to construct.

\pageenlarge{1} We construct both graphs using $k=8$ neighbours, and explore the robustness to various values of $k$ in Section~\ref{sec:robust}. Because the number of edges (i.e., neighbours) per node (i.e., passage) is known, the graphs are both stored as a \nic{uncompressed} sequence of docids. Using unsigned 32-bit integer docids, only 32 bytes per passage are needed, which amounts to 283 MB to store an MS~MARCO graph.\footnote{For context, the compressed document source is 1035MB, and the compressed PISA index of MS~MARCO is 647MB.} We note that there are likely approaches that reduce the computational overhead in graph construction by making use of approximate searches; we leave this for future work. \nic{The two graphs differ substantially in their content.\footnote{Only 3\% of passages agree on seven or eight neighbours across graphs, and 43\% of passages have no agreement on neighbours across graphs.}} \crc{We release these graphs through our implementation to aid other researchers and enable future works.}

\subsection{Other Parameters and Settings}

We use a \sys{} batch size of $b=16$ by default, matching a typical batch size for a neural cross-encoder model. We explore the robustness of \sys{} to \nic{various values of} $b$ in Section~\ref{sec:robust}. We explore two budgets: $c=100$ (a reasonable budget for a deployed re-ranking system, e.g.,~\cite{10.1145/3404835.3462889}) and $c=1000$ (the \textit{de facto} default threshold commonly used in shared tasks like TREC).

\section{Results and Analysis}\label{sec:res}

{
\renewcommand{\arraystretch}{0.8}
\begin{table*}
\centering
\caption{Effectiveness of \sys{} on TREC DL 2019 and 2020 in a variety of re-ranking pipelines and re-ranking budgets ($c$). The top result for each pipeline is in bold. Significant differences with the baseline (typical re-ranking) are marked with *, while insignificant differences are in grey (paired t-test, $p<0.05$, using Bonferroni correction).}
\label{tab:comp}
\begin{tabular}{l|rrr|rrr|rrr|rrr}
\toprule
\multicolumn{1}{l}{}&\multicolumn{3}{c}{DL19 (valid.) $c=100$}&\multicolumn{3}{c}{DL19 (valid.) $c=1000$}&\multicolumn{3}{c}{DL20 (test) $c=100$}&\multicolumn{3}{c}{DL20 (test) $c=1000$} \\
\cmidrule(lr){2-4}\cmidrule(lr){5-7}\cmidrule(lr){8-10}\cmidrule(lr){11-13}
Pipeline & nDCG & MAP & R@1k & nDCG & MAP & R@1k & nDCG & MAP & R@1k & nDCG & MAP & R@1k \\
\midrule

BM25>>MonoT5-base   &        0.665  &        0.417  &        0.755  &        0.699  &        0.483  &        0.755  &        0.672  &        0.421  &        0.805  &        0.711  &        0.498  &        0.805  \\
\idnt w/ \sys{BM25} &    *   0.697  &    *   0.456  &    *   0.786  & \gy{   0.727} & \gy{   0.490} &    *   0.827  &    *   0.695  & \gy{   0.439} &    *   0.823  &    *   0.743  & \gy{\bf0.501} &    *   0.874  \\
\idnt w/ \sys{TCT}  &    *\bf0.722  &    *\bf0.491  &    *\bf0.800  &    *\bf0.743  & \gy{\bf0.511} &    *\bf0.839  &    *\bf0.714  &    *\bf0.472  &    *\bf0.831  &    *\bf0.749  & \gy{\bf0.501} &    *\bf0.892  \\
\midrule
BM25>>MonoT5-3b     &        0.667  &        0.418  &        0.755  &        0.700  &        0.489  &        0.755  &        0.678  &        0.442  &        0.805  &        0.728  &        0.534  &        0.805  \\
\idnt w/ \sys{BM25} &    *   0.693  & \gy{   0.454} &    *   0.790  &    *   0.741  & \gy{   0.517} &    *   0.831  &    *   0.715  &    *   0.469  &    *   0.829  &    *   0.772  & \gy{   0.556} &    *   0.881  \\
\idnt w/ \sys{TCT}  &    *\bf0.715  &    *\bf0.484  &    *\bf0.806  &    *\bf0.746  & \gy{\bf0.522} &    *\bf0.846  &    *\bf0.735  &    *\bf0.512  &    *\bf0.837  &    *\bf0.787  &    *\bf0.564  &    *\bf0.899  \\
\midrule
BM25>>ColBERT       &        0.663  &        0.409  &        0.755  &        0.681  &        0.458  &        0.755  &        0.667  &        0.421  &        0.805  &        0.697  &        0.469  &        0.805  \\
\idnt w/ \sys{BM25} &    *   0.690  &    *   0.442  &    *   0.783  &    *   0.720  & \gy{   0.480} &    *   0.825  &    *   0.695  &    *   0.446  &    *   0.823  &    *   0.732  & \gy{   0.479} &    *   0.870  \\
\idnt w/ \sys{TCT}  &    *\bf0.716  &    *\bf0.475  &    *\bf0.798  &    *\bf0.727  & \gy{\bf0.482} &    *\bf0.841  &    *\bf0.707  &    *\bf0.463  &    *\bf0.829  &    *\bf0.740  & \gy{\bf0.481} &    *\bf0.887  \\
\midrule
\midrule
TCT>>MonoT5-base    &        0.708  &        0.472  &        0.830  &        0.704  &        0.473  &        0.830  &        0.698  &        0.488  &        0.848  &        0.693  &        0.471  &        0.848  \\
\idnt w/ \sys{BM25} &    *\bf0.728  & \gy{\bf0.484} & \gy{\bf0.852} &    *\bf0.733  & \gy{\bf0.480} &    *\bf0.883  &    *\bf0.719  &    *\bf0.501  & \gy{\bf0.861} &    *\bf0.719  & \gy{\bf0.473} &    *\bf0.881  \\
\idnt w/ \sys{TCT}  & \gy{   0.722} & \gy{   0.481} & \gy{   0.847} &    *   0.724  & \gy{   0.474} & \gy{   0.866} &    *   0.712  & \gy{   0.494} & \gy{   0.856} &    *   0.710  & \gy{   0.471} & \gy{   0.871} \\
\midrule
TCT>>MonoT5-3b      &        0.720  &        0.498  &        0.830  &        0.725  &        0.513  &        0.830  &        0.723  &        0.534  &        0.848  &        0.733  &        0.544  &        0.848  \\
\idnt w/ \sys{BM25} &    *\bf0.748  &    *\bf0.521  &    *\bf0.857  &    *\bf0.759  & \gy{\bf0.521} &    *\bf0.885  &    *\bf0.743  & \gy{\bf0.546} &    *\bf0.864  &    *\bf0.771  &    *\bf0.555  &    *\bf0.890  \\
\idnt w/ \sys{TCT}  &    *   0.742  &    *   0.517  & \gy{   0.849} &    *   0.749  & \gy{   0.516} &    *   0.868  &    *   0.741  &    *   0.545  &    *   0.861  &    *   0.759  & \gy{   0.551} &    *   0.880  \\
\midrule
TCT>>ColBERT        &        0.708  &        0.464  &        0.830  &        0.701  &        0.452  &        0.830  &        0.698  &        0.476  &        0.848  &        0.697  &        0.470  &        0.848  \\
\idnt w/ \sys{BM25} &    *\bf0.729  &    *\bf0.480  & \gy{\bf0.853} &    *\bf0.727  & \gy{\bf0.459} & \gy{\bf0.876} &    *\bf0.715  & \gy{\bf0.485} & \gy{\bf0.857} &    *\bf0.722  &    *\bf0.477  &    *\bf0.877  \\
\idnt w/ \sys{TCT}  &    *   0.722  & \gy{   0.474} & \gy{   0.845} &    *   0.715  & \gy{   0.452} & \gy{   0.852} &    *   0.711  &    *   0.484  &    *\bf0.857  &    *   0.713  & \gy{   0.473} & \gy{   0.864} \\
\midrule
\midrule
D2Q>>MonoT5-base    &        0.736  &        0.503  &        0.830  &        0.747  &        0.531  &        0.830  &        0.726  &        0.499  &        0.839  &        0.731  &     \bf0.508  &        0.839  \\
\idnt w/ \sys{BM25} &    *   0.748  & \gy{   0.506} & \gy{   0.848} & \gy{   0.757} & \gy{   0.519} &    *\bf0.880  &    *   0.734  & \gy{   0.497} &    *   0.847  & \gy{\bf0.748} & \gy{   0.504} &    *   0.880  \\
\idnt w/ \sys{TCT}  &    *\bf0.760  &    *\bf0.528  & \gy{\bf0.850} &    *\bf0.766  & \gy{\bf0.533} &    *   0.879  & \gy{\bf0.740} & \gy{\bf0.508} &    *\bf0.856  & \gy{\bf0.748} & \gy{   0.499} &    *\bf0.895  \\
\midrule
D2Q>>MonoT5-3b      &        0.737  &        0.506  &        0.830  &        0.751  &        0.542  &        0.830  &        0.738  &        0.531  &        0.839  &        0.753  &        0.557  &        0.839  \\
\idnt w/ \sys{BM25} & \gy{   0.744} & \gy{   0.512} &    *   0.850  & \gy{\bf0.772} & \gy{\bf0.549} &    *\bf0.880  &    *   0.751  & \gy{   0.535} &    *   0.852  &    *   0.781  & \gy{   0.561} &    *   0.887  \\
\idnt w/ \sys{TCT}  & \gy{\bf0.755} & \gy{\bf0.524} &    *\bf0.857  & \gy{   0.769} & \gy{   0.544} &    *\bf0.880  &    *\bf0.764  & \gy{\bf0.550} &    *\bf0.860  &    *\bf0.790  & \gy{\bf0.565} &    *\bf0.905  \\
\midrule
D2Q>>ColBERT        &        0.724  &        0.475  &        0.830  &        0.733  &        0.501  &        0.830  &        0.718  &        0.483  &        0.839  &        0.717  &        0.479  &        0.839  \\
\idnt w/ \sys{BM25} & \gy{   0.734} & \gy{   0.484} & \gy{   0.845} & \gy{\bf0.753} & \gy{\bf0.505} &    *   0.876  &    *   0.731  & \gy{   0.487} &    *   0.849  &    *   0.737  & \gy{   0.482} &    *   0.872  \\
\idnt w/ \sys{TCT}  &    *\bf0.744  &    *\bf0.496  & \gy{\bf0.849} &    *   0.752  & \gy{   0.503} &    *\bf0.878  &    *\bf0.735  & \gy{\bf0.488} &    *\bf0.856  &    *\bf0.746  & \gy{\bf0.485} &    *\bf0.893  \\
\midrule
\midrule
SPLADE>>MonoT5-base &        0.750  &        0.506  &        0.872  &        0.737  &     \bf0.487  &        0.872  &        0.748  &        0.505  &        0.899  &        0.731  &     \bf0.480  &        0.899  \\
\idnt w/ \sys{BM25} &    *\bf0.762  & \gy{   0.509} & \gy{\bf0.888} & \gy{\bf0.745} & \gy{\bf0.487} & \gy{\bf0.893} &    *\bf0.757  & \gy{\bf0.509} & \gy{   0.902} & \gy{\bf0.737} & \gy{   0.479} & \gy{\bf0.909} \\
\idnt w/ \sys{TCT}  &    *   0.759  & \gy{\bf0.512} & \gy{   0.878} & \gy{   0.737} & \gy{   0.481} & \gy{   0.875} & \gy{   0.751} & \gy{   0.506} & \gy{\bf0.903} & \gy{   0.734} & \gy{   0.475} & \gy{   0.908} \\
\midrule
SPLADE>>MonoT5-3b   &        0.761  &        0.526  &        0.872  &        0.764  &     \bf0.533  &        0.872  &        0.774  &        0.559  &        0.899  &        0.775  &        0.560  &        0.899  \\
\idnt w/ \sys{BM25} &    *\bf0.775  & \gy{   0.532} &    *\bf0.891  & \gy{\bf0.774} & \gy{\bf0.533} & \gy{\bf0.896} &    *\bf0.780  & \gy{   0.559} & \gy{   0.903} &    *\bf0.788  & \gy{\bf0.562} &    *\bf0.919  \\
\idnt w/ \sys{TCT}  &    *   0.773  & \gy{\bf0.539} & \gy{   0.884} & \gy{   0.769} & \gy{   0.531} & \gy{   0.881} &    *\bf0.780  & \gy{\bf0.561} & \gy{\bf0.905} & \gy{   0.783} & \gy{   0.559} & \gy{   0.910} \\
\midrule
SPLADE>>ColBERT     &        0.741  &        0.479  &        0.872  &        0.727  &     \bf0.456  &        0.872  &        0.747  &        0.495  &        0.899  &        0.733  &        0.474  &        0.899  \\
\idnt w/ \sys{BM25} &    *\bf0.753  & \gy{\bf0.490} & \gy{\bf0.885} & \gy{\bf0.730} & \gy{\bf0.456} & \gy{\bf0.875} &    *\bf0.755  & \gy{\bf0.501} & \gy{   0.902} &    *\bf0.742  &    *\bf0.477  & \gy{\bf0.914} \\
\idnt w/ \sys{TCT}  &    *   0.750  & \gy{   0.489} & \gy{   0.876} & \gy{   0.727} & \gy{   0.455} & \gy{   0.868} &    *   0.752  & \gy{   0.500} & \gy{\bf0.903} & \gy{   0.740} &    *   0.476  & \gy{   0.911} \\

\bottomrule
\end{tabular}
\end{table*}
}

\begin{table}
\centering
\caption{Effectiveness of \sys{} on the MS~MARCO dev (small) set with a re-ranking budget of $c=100$. The top result for each pipeline is in bold. Significant differences with the baseline (typical re-ranking) are marked with * (paired t-test, $p<0.05$, using Bonferroni correction).}\vspace{-0.5em}
\label{tab:dev}
\begin{tabular}{lrrrr}
\toprule
&\multicolumn{2}{c}{>>MonoT5-base}&\multicolumn{2}{c}{>>ColBERT}\\
\cmidrule(lr){2-3}\cmidrule(lr){4-5}
Pipeline & RR@10 & R@1k & RR@10 & R@1k \\
\midrule
BM25>>           &     0.356 &     0.868  &     0.323 &     0.868 \\
\idnt \sys{BM25} &     0.358 & *   0.881  &     0.323 & *   0.882 \\
\idnt \sys{TCT}  & *\bf0.369 & *\bf0.903  & *\bf0.333 & *\bf0.902 \\
\midrule
TCT>>            &     0.388 &     0.970  &     0.345 &     0.970 \\  
\idnt \sys{BM25} &  \bf0.389 & *\bf0.973  & *\bf0.346 & *\bf0.973 \\  
\idnt \sys{TCT}  &     0.388 & *\bf0.973  &  \bf0.346 & *   0.972 \\
\midrule
D2Q>>            &  \bf0.386 &     0.936  &  \bf0.345 &     0.936 \\
\idnt \sys{BM25} &  \bf0.386 & *   0.941  &  \bf0.345 & *   0.941 \\
\idnt \sys{TCT}  &  \bf0.386 & *\bf0.949  &     0.344 & *\bf0.948 \\
\midrule
SPLADE>>         &  \bf0.389 &     0.983  &     0.345 &     0.983 \\
\idnt \sys{BM25} &  \bf0.389 &  \bf0.984  & *\bf0.346 &  \bf0.984 \\
\idnt \sys{TCT}  &     0.388 & *\bf0.984  & *\bf0.346 &  \bf0.984 \\
\bottomrule
\end{tabular}\vspace{-1em}
\end{table}

We now present the results of our experiments and conduct associated analysis to answer our research questions.

\subsection{Effectiveness}\label{sec:effect}

\looseness -1 To understand whether \sys{} is generally effective, it is necessary to test the effect it has on a variety of retrieval pipelines. Therefore, we construct re-ranking pipelines based on every pair of our initial ranking functions (BM25, TCT, D2Q, and SPLADE) and scoring functions (MonoT5-base, MonoT5-3b, and ColBERT). These \nic{12} pipelines collectively cover a variety of paradigms. Table~\ref{tab:comp} presents the results of \sys{} on these pipelines for TREC DL 2019 and 2020 using both the lexical BM25-based graph and the semantic TCT-based corpus graph. \nic{We report results using both re-ranking budgets $c=100$ and $c=1000$.}

\pageenlarge{1} Each box in Table~\ref{tab:comp} allows the reader to \nic{inspect} the effect on retrieval effectiveness that \sys{} has on a particular re-ranking pipeline and re-ranking budget. In general, we see that the greatest improvement when the initial retrieval pool is poor. In particular, BM25 only provides a R@1k of 0.755 and 0.805 on DL19 and DL20, respectively, while improved retrieval functions offer up to 0.872 and 0.899, \nic{respectively} (SPLADE). \sys{} enables the pipelines to find additional relevant documents. Using BM25 as \nic{the} initial \nic{pool}, our approach reaches a R@1k up to 0.846 and 0.892, respectively (BM25>>MonoT5-3b w/ \sys{TCT} \nic{ and $c=1000$}). Perhaps unsurprisingly, this result is achieved using both a corpus graph \nic{(\sys{TCT})} that differs substantially from the technique used for initial retrieval \nic{(BM25)} \nic{and using the most effective re-ranking function (MonoT5-3b)}. However, we also note surprisingly high recall in this setting when using the \sys{BM25} corpus graph: up to 0.831 (DL19) and 0.881 (DL20). These results are on par with the recall achieved by TCT and D2Q -- an impressive feat considering that this pipeline only uses lexical signals and a single neural model trained with a conventional process.\footnote{D2Q is trained as a sequence-to-sequence model and involves a lengthy inference stage during indexing, while TCT employs a complex, multi-stage training process involving another trained scoring model \crc{that is challenging to fully reproduce~\cite{wang:sigir2022-tctrepro}}. Meanwhile, MonoT5-3b is simply trained using MS~MARCO's training triples.} The pipelines that use a BM25 initial ranker also benefit greatly in terms of nDCG, which is likely \nic{due} in part to the improved recall.

\looseness -1 Significant improvements are also observed in all other pipelines, particularly in terms of nDCG when there is a low re-ranking budget available ($c=100$) and in recall when a high budget is available ($c=1000$). In general, the corpus graph that is least \nic{similar to} the initial ranker is most effective (e.g., the BM25 graph when using a TCT ranking). However, we note that both corpus graphs improve every pipeline, at least in some settings. For instance, the \sys{TCT} corpus graph consistently improves the nDCG of pipelines that use TCT as an initial ranker, but rarely the recall.

We also note that \sys{} can nearly always \nic{improve} the precision of the top results, as measured by nDCG, in settings with a limited re-ranking budget ($c=100$), even when R@1k remains unchanged. This is likely due to the fact that \sys{} is able to pick out documents from lower depths of the initial ranking pool to score within the limited available budget. For instance, \nic{in the case of} the \crc{strong} SPLADE>>MonoT5-base pipeline \nic{with $c=100$}, which offers high recall to begin with \nic{(0.872 on DL19 and 0.899 on DL20)}, \sys{BM25} improves the nDCG from 0.750 to 0.762 (DL19) and from 0.748 to 0.757 (DL20), while leaving the R@1k unchanged.

\looseness -1 In a few rare cases, we observe that \sys{} can yield a lower mean performance than the baseline (e.g., MAP for the D2Q>>MonoT5-base pipeline with $c=1000$). However, these differences are never statistically significant and are usually accompanied by significant improvements to other measures (e.g., the R@1k improves).

We note that the same trends appear for both our validation set (DL19) and our held-out test set (DL20), suggesting that \sys{} is not over-fit\nic{ted} to the data that we used during the development of \sys{}.

\looseness -1 Finally, we test \sys{} on the MS~MARCO dev (small) set. This setting differs from the TREC DL experiments in that each of the queries has only a few (usually just one) passages that are labeled as relevant, but has far more queries (6,980 compared to 43 in DL19 and 54 in DL20). Thus, experiments on this dataset test a pipeline's capacity to retrieve a single (and somewhat arbitrary) relevant passage for a query.\footnote{The suitability of this dataset \crc{for evaluation} is debated in the community (e.g.,~\cite{https://doi.org/10.48550/arxiv.2109.00062,https://doi.org/10.48550/arxiv.2112.03396}), but we include it for completeness.} Due to the \nic{cost} of running multiple versions of highly-expensive re-ranking pipelines, we limit this study to a low re-ranking budget $c=100$ and to the two less expensive scoring functions (MonoT5-base and ColBERT). Table~\ref{tab:dev} presents the results. We find that \sys{} offers the most benefit in pipelines that suffer from the lower recall -- namely, the BM25-based pipelines. In this setting, the improved R@1k also boosts the RR@10. In the TCT, D2Q, and SPLADE pipelines, R@1k often significantly improved, but this results in non-significant (or marginal) changes to RR@10.

To answer RQ1, we find that \sys{} provides significant benefits in terms of precision- and recall-oriented measures. The results hold across a variety of initial retrieval functions, re-ranking functions, and re-ranking budgets. The most benefit is apparent when the initial pool has low recall, though we note that \sys{} also improves over systems with high initial recall -- particularly by enabling higher precision at a lower re-ranking budget. \crc{Overall, we find that \sys{} is safe to apply to any re-ranking pipeline (i.e., it will not harm the effectiveness), and it will often improve performance (particularly when the re-ranking budget is limited or when a low-cost first stage retriever is used).}

\begin{figure}
\centering
\includegraphics[scale=0.55]{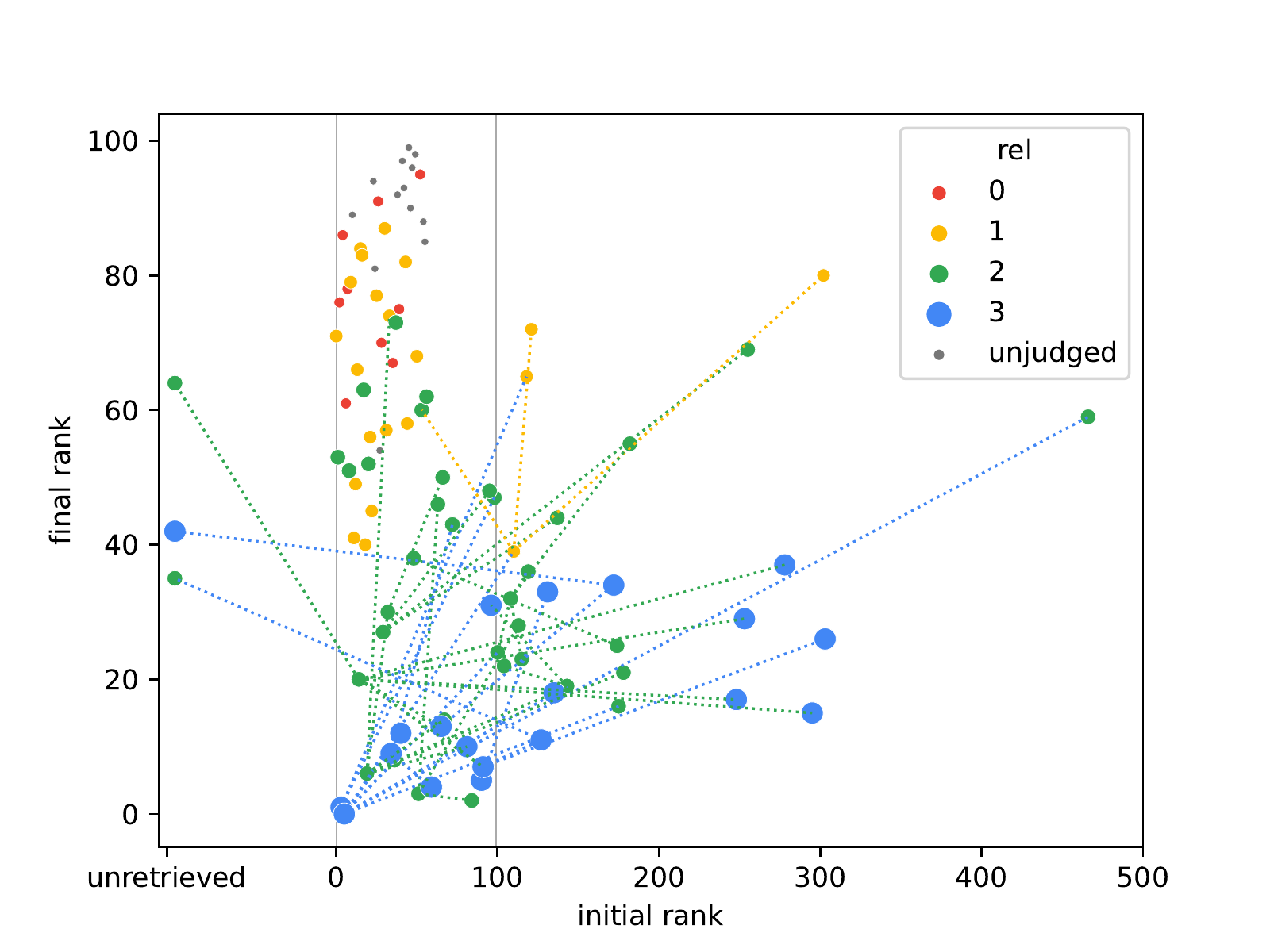}\vspace{-1.5em}
\caption{Plot of the initial and final rankings of BM25>>MonoT5-base using \sys{TCT} with $c=100$ for the DL19 query \textit{`how long is life cycle of flea'}. The colour/size of dots indicate the relevance \nic{label}. Lines between points indicate links followed in the corpus graph.}
\label{fig:neato}\vspace{-1em}
\end{figure}

\looseness -1 \nic{To illustrate the ability of \sys{} to promote low-ranked documents under limited ranking budgets, Figure~\ref{fig:neato} plots the initial rank (x-axis) of documents and their final rank (y-axis), for a particular query. Each point represents a retrieved document, with colour/size indicative of the relevance label. Lines between points indicate links followed in the corpus graph. It can be seen that by leveraging the corpus graph, \sys{} is able to promote highly relevant documents that were lowly scored in the initial ranking, as well as retrieve `new' relevant documents, \crc{which} are not retrieved in the initial BM25 pool.} \crc{For instance, GAR is able to select five rel=2 documents from around initial rank 250-300, and ultimately score them within the top 40 documents. Meanwhile, it retrieves two rel=2 and one rel=3 documents that were not found in the first stage.}

\subsection{Computational Overhead}\label{sec:perf}

\sys{} is designed to have a minimal impact on query latency. By relying on a pre-computed corpus graph that will often be small enough to fit into memory (283MB with $k=8$ for MS~MARCO), neighbour lookups are performed in $O(1)$ time. With the frontier $F$ stored in a heap, insertions take only $O(1)$, meaning that finding neighbours and updating the frontier adds only a constant time \nic{for each} scored document. Sampling the top $b$ items from the heap takes $O(b \log c)$, since the number of items in the heap never needs to exceed the budget $c$.

\looseness -1 To obtain a practical sense of the computational overhead of \sys{}, we conduct latency tests. To isolate the effect of \sys{} itself, we find it necessary to factor out the overhead from the re-ranking model itself, since the variance in latency between neural scoring runs often exceeds the overhead introduced by \sys{}. To this end, we pre-compute and store all the needed query-document scores and simply look them up as they would be scored. We then test various re-ranking budgets ($c$) for DL19, and take 10 latency measurements of the typical re-ranking and \sys{} processes. Table~\ref{tab:latency} reports the differences between the latency of \sys{} and the typical re-ranking results, isolating the overhead of \sys{} itself. We find that \sys{} introduces less than \nic{37.37ms overhead per 1000 documents scores (i.e., 2.68-3.73ms overhead per 100 documents scored)}, on average, \nic{using $16$ documents per batch}. We report results using the semantic TCT-based corpus graph, though we find little difference when using the lexical BM25-based corpus graph. The overhead can be further reduced (down to \nic{3.1ms} per 100 documents) by using a larger batch size\nic{, i.e., $64$ documents per batch}; we explore the effect of the batch size parameter on effectiveness in Section~\ref{sec:robust}. When compared to the cost of monoT5 scoring \nic{(rightmost column in Table~\ref{tab:latency})}, the \sys{} process adds negligible overhead, typically amounting to less than a \nic{2}\% increase in latency and falls within the variance of the scoring function's latency for low re-ranking budgets.

This experiment answers RQ2: the online computational overhead of \sys{} is minimal. It can be efficiently implemented using a heap, and adds only around \nic{3-4ms} per 100 documents in the re-ranking budget. This overhead is negligible when compared with the latency of a leading neural scoring function, though it will represent a higher proportion for more efficient scoring functions.

\begin{table}[b]
\centering
\caption{Mean latency overheads (ms/query) for \sys{} with 95\% confidence intervals. The latency of MonoT5-base scoring (with a model batch size of 64) is presented for context.}\vspace{-0.5em}
\label{tab:latency}
\begin{tabular}{rrrr}
\toprule
&\multicolumn{2}{c}{\sys{TCT}} & \multicolumn{1}{c}{MonoT5-base} \\
\cmidrule(lr){2-3}
c &  $b=16$ & $b=64$ & \multicolumn{1}{c}{Scoring} \\
\midrule
100        &  $2.68\pm0.02$ &  $0.57\pm0.01$ &   $267.06\pm6.12$ \\
250        &  $8.10\pm0.05$ &  $4.34\pm0.01$ &   $652.30\pm7.53$ \\
500        & $17.38\pm0.07$ & $13.66\pm0.02$ & $1,362.14\pm5.27$ \\
750        & $26.96\pm0.12$ & $22.29\pm0.07$ & $2,047.20\pm6.71$ \\
1000       & $37.37\pm0.07$ & $30.82\pm0.04$ & $2,631.75\pm6.28$ \\
\bottomrule
\end{tabular}\vspace{-1em}
\end{table}

\subsection{Robustness to Parameters}\label{sec:robust}

\begin{figure}
\centering
\def\figscale{0.59}
\includegraphics[scale=\figscale]{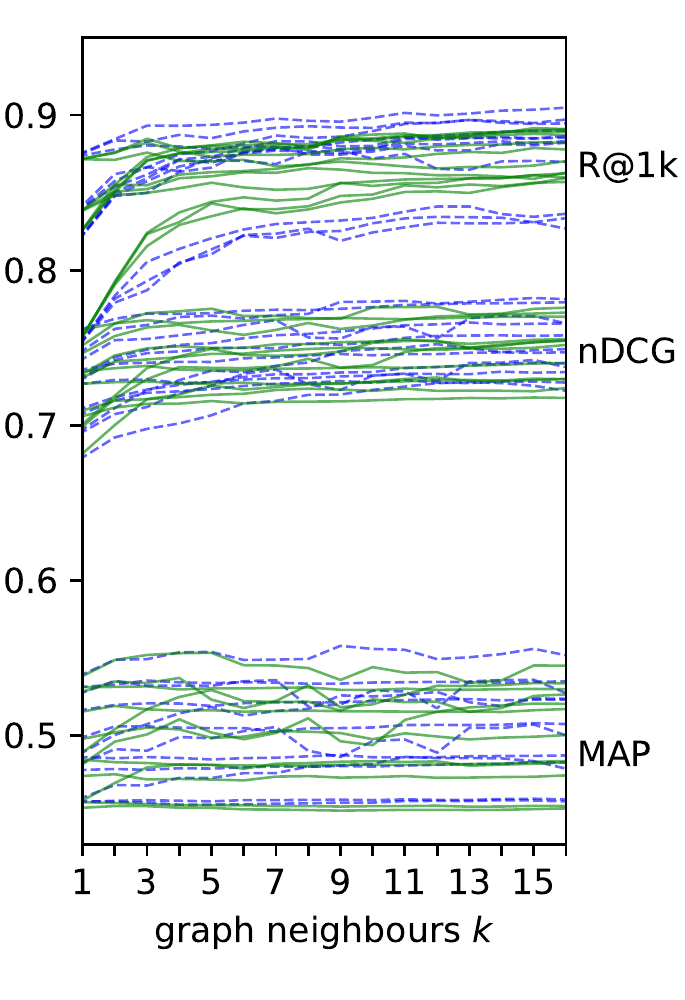}
\includegraphics[scale=\figscale]{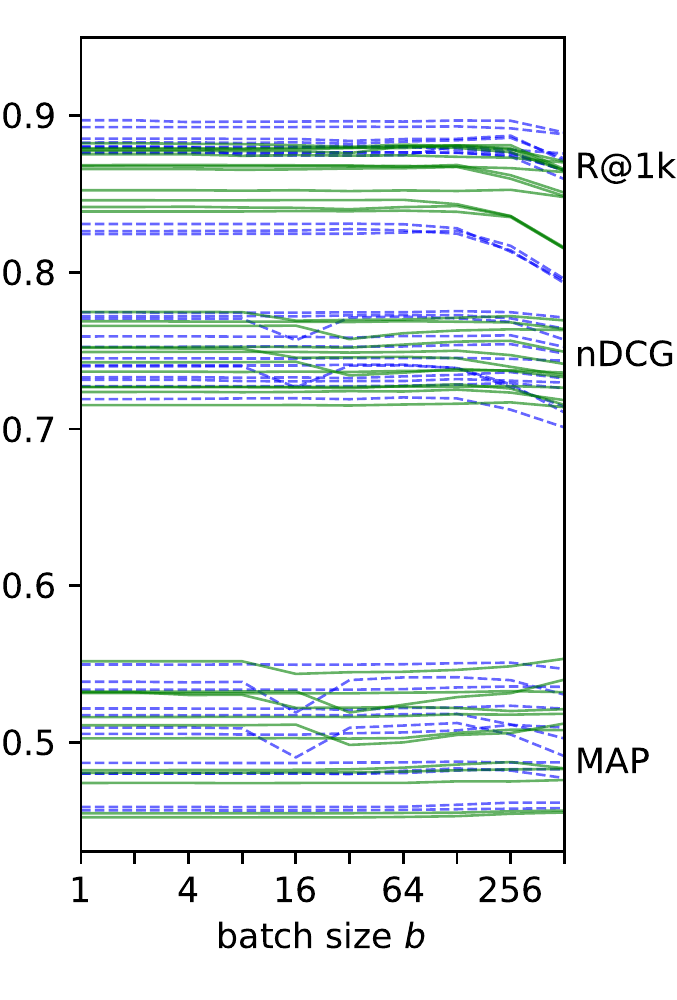}\vspace{-1em}
\caption{Performance of \sys{} when the number of neighbours in the corpus graph $k$ and the batch size $b$ vary. Each line represents a system from Table~\ref{tab:comp}. The dashed blue (solid green) lines are for the BM25 (TCT) graph.}\vspace{-1em}
\label{fig:robust}
\end{figure}

\looseness -1 Recall that \sys{} introduces two new parameters: the number of nearest neighbours in the corpus graph $k$ and the batch size $b$. In this section, we conduct experiments to test whether \sys{} is robust to the settings of these parameters.\footnote{Due to the number of pipelines and parameter settings, an exhaustive grid search over these parameters is prohibitively expensive.} We separately sweep $k\in[1,16]$ and $b\in[1,512]$ (by powers of 2) over DL19 with $c=1000$ for all \sys{} pipelines\nic{, and present the different effectiveness metrics in Figure~\ref{fig:robust}}.

With regard to the number of graph neighbours $k$, \nic{the nDCG, MAP and recall metrics are relatively stable from around $k=6$ to $k=16$ for almost all pipelines}. The MAP performance appears to be the least stable in this range, with some fluctuations in performance between $k=7$ and $k=13$. Recall appears to be most affected, with sharp gains for some pipelines between $k=1$  to $k=4$. \nic{This trend is present also for nDCG.}

The batch size $b$ is remarkably stable from $b=1$ to $b=128$, with only a blip in effectiveness for the BM25 graph at $b=16$. The most prominent shift in performance occurs at large batch sizes, e.g., $b=512$. We note that, \nic{when $b=512$}, the corpus graph can only be traversed for a single hop -- the neighbours of the top-scoring documents from the frontier batch are not able to be fed back into the re-ranking pool. This validates our technique of incorporating the feedback mechanism into the re-ranking process itself, which gives the model more chances to traverse the graph. While it may be tempting to prefer the stability of the system with very low batch sizes, we note that this has an effect on the performance: as seen in Section~\ref{sec:perf}, lower batch sizes reduces the speed of \sys{} itself. Further, and more importantly, $b$ imposes a maximum batch size of the scoring function itself; given that neural models benefit considerably in terms of performance with larger batch sizes (since the operations on the GPU are parallelised), larger values of $b$ (e.g., $b=16$ to $b=128$) should be preferred for practical reasons.

To answer RQ3, \crc{we find that} the performance of \sys{} is stable across various pipelines when the number of neighbours is sufficiently large ($k\geq6$) and the batch size is sufficiently low ($b\leq128$).

\subsection{Baseline Performance}\label{sec:horserace}

{
\rowcolors{5}{}{lightgray}
\begin{table*}
\centering
\caption{Performance of \sys, compared to a variety of other baselines. Significant differences are computed within \nic{groups}, with significance denoted as superscript letters $^{a-c}$ (paired t-test, $p<0.05$, Bonferroni correction). Rows marked with $\dagger$ are given to provide additional context, but the metrics were copied from other papers so do not include statistical tests.}\vspace{-0.5em}
\label{tab:horserace}
\resizebox{176mm}{!}{
\begin{tabular}{cllllrrrrrrrr}
\toprule
&&&&&\multicolumn{4}{c}{DL19 (validation)}&\multicolumn{4}{c}{DL20 (test)} \\
\cmidrule(lr){6-9}\cmidrule(lr){10-13}
&$R_0$&$R_1$&$R_2$& RR & nDCG@10 & nDCG & R@1k & Judged@10 & nDCG@10 & nDCG & R@1k & Judged@10 \\
\midrule

\multicolumn{11}{l}{\bf Kitchen Sink Systems} \\
           & D2Q         & >>MonoT5-3b   & >>DuoT5-3b  &                                     & \bf0.771 &    0.756 & $^{ab}$0.830 & 0.958 &    0.785 & $^{ab}$0.754 & $^{ab}$0.839 & 0.996 \\
           & SPLADE      & >>MonoT5-3b   & >>DuoT5-3b  &                                     &    0.768 &    0.772 &        0.872 & 0.953 &    0.787 &  $^{a}$0.781 &  $^{a}$0.899 & 0.987 \\
a          & SPLADE      & >>MonoT5-3b   & >>DuoT5-3b  & \sys{BM25}                          &    0.767 & \bf0.781 &     \bf0.896 & 0.951 &    0.787 &     \bf0.794 &     \bf0.919 & 0.989 \\
b          & D2Q         & >>MonoT5-3b   & >>DuoT5-3b  & \sys{TCT}                           &    0.766 &    0.775 &        0.880 & 0.953 & \bf0.788 &        0.793 &        0.905 & 0.993 \\
 $\dagger$ & TAS-B+D2Q~\cite{Hofsttter2021EfficientlyTA}   & >>MonoT5-3b   & >>DuoT5-3b  &  &    0.759 &        - &        0.882 &     - &    0.783 &            - &        0.895 &     - \\

\midrule
\multicolumn{11}{l}{\bf Single-Model Systems} \\
           & ColBERT ANN & >>ColBERT     &>>ColBERT-PRF&             &    \bf0.739 &    \bf0.764 &       0.871 & 0.907 &       0.715 &       0.746 &       0.884 & 0.946 \\
           & SPLADE      & -             & -           &             &       0.731 &       0.755 &    \bf0.872 & 0.926 &       0.720 &       0.750 &    \bf0.899 & 0.970 \\
c          & BM25        & >>MonoT5-3b   & -           & \sys{BM25}  &       0.729 &       0.741 &       0.831 & 0.947 &    \bf0.756 &    \bf0.772 &       0.881 & 0.972 \\
           & BM25        & >>MonoT5-3b   & -           &             &       0.722 &       0.700 & $^{c}$0.755 & 0.944 &       0.749 & $^{c}$0.728 & $^{c}$0.805 & 0.980 \\
           & TCT         & -             & -           &             &       0.721 &       0.708 &       0.830 & 0.914 & $^{c}$0.686 & $^{c}$0.689 &       0.848 & 0.931 \\
           & ColBERT ANN & >>ColBERT     & -           &             &       0.693 &       0.687 &       0.789 & 0.884 & $^{c}$0.687 & $^{c}$0.711 &       0.825 & 0.937 \\
 $\dagger$ & ANCE        & -             & -           &             &       0.648 &           - &       0.755 & 0.851 &       0.646 &           - &       0.776 & 0.865 \\
           & D2Q         & -             & -           &             & $^{c}$0.615 & $^{c}$0.678 &       0.830 & 0.916 & $^{c}$0.608 & $^{c}$0.676 &       0.839 & 0.956 \\

\bottomrule
\end{tabular}}
\end{table*}
}

\looseness -1 \crc{Section~\ref{sec:effect} established the effectiveness of \sys{} as ablations over a variety of re-ranking pipelines. We now explore how the approach fits into the broader context of the approaches proposed for passage retrieval and ranking. We explore two classes of pipelines: `Kitchen Sink' approaches that combine numerous approaches and models together, and `Single-Model' approaches that use only involve a single neural model at any stage.} We select representative \sys{} variants based on the nDCG@10 performance on DL19 (i.e., as a validation set), with DL20 again treated as the held-out test set. All systems use a re-ranking budget of $c=1000$. In this table, we report nDCG@10 to allow comparisons against prior work. We also report the judgment rate at 10 to provide context about how \nic{missing information} in the judgments may affect the nDCG@10 scores.

\crc{The Kitchen Sink results are reported in the top section of Table~\ref{tab:horserace}. All systems involve three ranking components: an initial retriever $R_0$, a Mono-scorer $R_1$ (which assigns a relevance score to each document), and a Duo-scorer $R_2$ (which scores and aggregates pairs of documents). The Duo-style models are known to improve the ranking of the top documents~\cite{10.48550/arxiv.2101.05667}.} Although we leave the exploration of how \sys{} can be used to augment the Duo process directly for future work, we still want to check what effect \sys{} has on these pipelines. We ablate two Duo systems (either based on D2Q or SPLADE) using \sys{} for the first-stage re-ranker and a DuoT5-3b-based second-stage re-ranker (second stage uses the suggested cutoff of 50 from~\cite{10.48550/arxiv.2101.05667}). We observe that there is no significant difference in terms of precision of the top 10 results. \crc{However, \sys{} can still provide a significant improvement in terms of nDCG later in the ranking and in terms of recall.} These results suggest that although \sys{} identifies more relevant documents, the Duo models are not capable of promoting them to the top ranks.

\crc{We next explore Single-Model systems, which are shown in the bottom section of Table~\ref{tab:horserace}.} Having only a single models likely has some practical advantages: pipelines that use a single model tend to be simpler, and practitioners only need to train a single model. Here, we compare with a variety of systems that fall into this category, most notably the recently-proposed ColBERT-PRF approaches that operate over dense indexes~\cite{10.1145/3471158.3472250}. A \sys{BM25} pipeline that operates over BM25 results also falls into this category, since only a single neural model (the scorer) is needed. Among this group, \sys{} performs competitively, outmatched only by ColBERT-PRF~\cite{10.1145/3471158.3472250} and the recent SPLADE~\cite{https://doi.org/10.48550/arxiv.2205.04733} model (though the differences in performance are not statistically significant). Compared to these methods, though, \sys{} requires far less storage -- the corpus graph for \sys{} is only around 283MB, while the index for SPLADE is 8GB, and the vectors required for ColBERT-PRF are 160GB.

{
\renewcommand{\arraystretch}{0.8}
\begin{table}
\centering
\caption{Intra-List Similarity (ILS) among retrieved relevant documents. Since the set of retrieved documents does not change using typical Re-Ranking (RR), each value in this column is only listed once. ILS scores that are statistically equivalent to the RR setting are indicated with * (procedure described in Section~\ref{sec:divers}).}\vspace{-0.5em}
\label{tab:divers}
\resizebox{85mm}{!}{
\begin{tabular}{lrrrrr}
\toprule
&& \multicolumn{2}{c}{\sys{BM25}}& \multicolumn{2}{c}{\sys{TCT}} \\
\cmidrule(lr){3-4}\cmidrule(lr){5-6}
Pipeline & RR & $c$=100 & $c$=1k & $c$=100 & $c$=1k \\
\midrule

BM25>>MonoT5-base  & 0.947 & * 0.946 & * 0.946 & * 0.947 & * 0.946 \\
BM25>>MonoT5-3b    &       & * 0.946 & * 0.946 & * 0.946 & * 0.946 \\
BM25>>ColBERT      &       & * 0.946 & * 0.946 & * 0.947 & * 0.946 \\
\midrule
TCT>>MonoT5-base   & 0.969 & * 0.969 & * 0.968 & * 0.969 & * 0.969 \\
TCT>>MonoT5-3b     &       & * 0.969 & * 0.968 & * 0.969 & * 0.969 \\
TCT>>ColBERT       &       & * 0.969 & * 0.969 & * 0.969 & * 0.969 \\
\midrule
D2Q>>MonoT5-base   & 0.969 & * 0.968 & * 0.968 & * 0.969 & * 0.968 \\
D2Q>>MonoT5-3b     &       & * 0.968 & * 0.968 & * 0.968 & * 0.968 \\
D2Q>>ColBERT       &       & * 0.968 & * 0.968 & * 0.969 & * 0.968 \\
\midrule
SPLADE>>MonoT5-base& 0.969 & * 0.968 & * 0.968 & * 0.969 & * 0.969 \\
SPLADE>>MonoT5-3b  &       & * 0.968 & * 0.968 & * 0.968 & * 0.969 \\
SPLADE>>ColBERT    &       & * 0.968 & * 0.969 & * 0.969 & * 0.969 \\
\bottomrule
\end{tabular}}\vspace{-1em}
\end{table}
}

To answer RQ4: We observe that \sys{} can be incorporated into a variety of larger, state-of-the-art re-ranking pipelines. It frequently boosts the recall of systems that it is applied to, though the scoring functions we explore tend to have difficulty in making use of the additional relevant passages. This motivates exploring further improvements to re-ranking models. For instance, cross-encoder models have largely relied on simple BM25 negative sampling (e.g., from the MS~MARCO triples file) for training. Techniques like hard negative sampling~\cite{ance} and distillation~\cite{lin-etal-2021-batch} (employed to train models like SPLADE and TCT) have so far been largely unexplored for cross-encoder models; these techniques may help them recognise more relevant documents.

\subsection{Diversity of Retrieved Passages}\label{sec:divers}

Next, we test whether \sys{} results in a more homogeneous set of retrieved relevant passages, compared to typical re-ranking. Among the set of relevant passages each system retrieved,\footnote{We are only concerned with the diversity among the relevant passages ($rel=2$ or $rel=3$) because non-relevant passages are inherently dissimilar from relevant ones.} we compute the Intra-List Similarity (ILS)~\cite{ziegler-2005-improving} using our TCT embeddings. \crc{ILS is the average cosine similarity between all pairs of items in a set, so a higher ILS values here indicate that the relevant documents are more similar to one another.} Table~\ref{tab:divers} compares the ILS of each initial ranking function (BM25, TCT, D2Q, and SPLADE) with the \sys{BM25} and \sys{TCT} counterparts. Using two-one-sided t-tests (TOSTs) with bounds of 0.005 and $p<0.05$ (including a Bonferonni correction), we find that \sys{} yields statistically equivalent diversity to the typical re-ranking system.

These results answer RQ5: despite using document similarity to help choose additional documents to score, \sys{} does not result in the system retrieving a more homogeneous set of relevant passages.

\section{Conclusions and Outlook}\label{sec:concl}

In this paper we took a modern approach to the cluster hypothesis, to consider nearest neighbour documents  while re-ranking using a neural re-ranker. In this way, the neural re-ranker feedbacks how useful documents are, which allows to prioritise further the next batch of documents to identify in an adaptive manner. Experiments using nearest neighbour graphs computed using BM25 and TCT-ColBERT-HNP demonstrated the promise of our Graph-based Adaptive Re-ranking approach, with significant improvements in precision- and recall-oriented measures on the TREC DL 2019 \& 2020 passage ranking corpus. For instance, \nic{\sys{} can improve nDCG by up to 8\% (BM25>>monoT5-base w/ TCT) and R@1000 up to 12\% (also BM25>>monoT5-base w/ TCT)}. 

We believe this work opens up several \nic{directions for the modelling of adaptive re-ranking} - indeed, it can be seen as a {\em closed loop} modelling problem (as opposed to the classical {\em open loop} re-ranking formulation), or as an explore/exploit scenario. Due to the effectiveness and efficiency of the instantiations shown here, we leave further advanced formulations to future work. 

\section*{Acknowledgements}
Sean and Craig acknowledge EPSRC grant EP/R018634/1: Closed-Loop Data Science for Complex, Computationally- \& Data-Intensive Analytics.

\newcounter{BalanceAtReference}
\setcounter{BalanceAtReference}{19}
\newcounter{ReferenceIndexForBalancing}

\makeatletter

\global\@ACM@balancefalse

\def\@balancelastpageonce{%
  \ifnum\value{ReferenceIndexForBalancing}=\value{BalanceAtReference}
    \newpage
  \else
    \relax
  \fi
  \stepcounter{ReferenceIndexForBalancing}
}
\pretocmd{\bibitem}{\@balancelastpageonce}
  {} %
  {\@latex@error{Patching \bibitem failed}{\@ehd}}

\makeatother

\bibliographystyle{ACM-Reference-Format}

\bibliography{biblio}%
\balance

\end{document}